\documentstyle[emulateapj]{article}

\submitted{Accepted 1998 May 1; Astrophysical Journal Letters}

\begin{document}

\title{Silicon Nanoparticles: Source of Extended Red Emission?}

\author{Adolf N.\ Witt}
\affil{Department of Physics \& Astronomy, The University of Toledo, 
  Toledo, OH 43606}
\author{Karl D.\ Gordon}
\affil{Department of Physics \& Astronomy, Louisiana State University,
  Baton Rouge, LA 70803}
\and
\author{Douglas G.\ Furton}
\affil{Department of Physical Sciences, Rhode Island College, 
   Providence, RI 02908}

\lefthead{Witt, Gordon, \& Furton}
\righthead{ERE from Si nanoparticles?}

\begin{abstract}
  We have reviewed the characteristics of the extended red emission
(ERE) as observed in many dusty astronomical environments, in
particular, the diffuse interstellar medium of the Galaxy. The
spectral nature and the photon conversion efficiency of the ERE
identify the underlying process as highly efficient photoluminescence
by an abundant component of interstellar dust. We have compared the
photoluminescence properties of a variety of carbon- and silicon-based
materials proposed as sources for the ERE with the observationally
established constraints. We found that silicon nanoparticles provide
the best match to the spectrum and the efficiency requirement of the
ERE. If present in interstellar space with an abundance sufficient to
explain the intensity of the ERE, silicon nanoparticles will also
contribute to the interstellar 9.7~$\micron$ Si-O stretch feature in
absorption, to the near- and mid-IR nonequilibrium thermal background
radiation, and to the continuum extinction in the near- and far-UV.
About 36\% of the interstellar silicon depleted into the dust phase
would be needed in the form of silicon nanoparticles, amounting to
less than 5\% of the interstellar dust mass. We propose that silicon
nanoparticles form through the nucleation of SiO in oxygen-rich
stellar mass outflows and that they represent an important small-grain
component of the interstellar dust spectrum.
\end{abstract}


\section{Introduction}

  The recent detection of extended red emission (ERE) in the diffuse
interstellar medium (ISM) of the Galaxy by Gordon et al.\ (1998) and
the spectroscopic confirmation of its luminescence band nature by
Szomoru \& Guhathakurta (1998) fundamentally changed our perception of
the ERE as an interstellar process in several ways. One, the ERE is no
longer seen as a phenomenon limited to localized special environments,
such as reflection and planetary nebulae. As a Galaxy-wide process,
its characteristics must now be imposed as an additional observational
constraint on models for interstellar dust in the diffuse ISM. Two, by
correlating the ERE intensity with the \ion{H}{1} column density in
the diffuse ISM at intermediate and high Galactic latitudes, and thus
with the dust column density, it has become possible to determine the
efficiency with which UV/visible photons of the interstellar radiation
field absorbed by interstellar dust are converted into ERE photons. An
ERE photon conversion efficiency of ($10 \pm 3$)\% (Gordon et al.\
1998; Szomoru \& Guhathakurta 1998) was derived by assuming the the
ERE agent consumes {\em all} photons absorbed by dust in the
91.2-550~nm wavelength range. To the extent that we know of the
existence of other dust components which are not expected to
contribute to the ERE, this assumption is false. The ERE agent, most
likely, is not responsible for the absorption of all absorbed
UV/visible photons. Thus, the derived efficiency is only a lower limit
to the true, intrinsic, efficiency of the photoluminescence (PL)
process seen as ERE. However, even if the intrinsic efficiency is as
high as 50\%, a reasonable upper limit for naturally occurring PL, the
ERE agent still needs to absorb about 20\% of all absorbed UV/visible
photons in the diffuse ISM. This can only be, if the ERE agent
consists of a cosmically abundant material which is capable of highly
efficient PL. It appears that only carbon- or silicon-based materials
fit these criteria. In this letter, we present the case for silicon
nanoparticles being the component of the ISM responsible for the ERE.

\section{Characteristics of the ERE and the implied properties of its carrier}

  We briefly summarize the observed characteristics of the ERE and
refer to the original sources for the details of the observations.

1.\ ERE manifests itself through a broad, featureless emission band of
60 $<$ FWHM $<$ 100 nm, with a peak appearing in the general
wavelength range $610 < \lambda_p < 820$~nm. The presence of ERE has
been established spectroscopically in many dusty astronomical
environments, e.g.\ the diffuse ISM (Szomoru \& Guhathakurta 1998),
reflection nebulae (Witt \& Boroson 1990), planetary nebulae (Furton
\& Witt 1990,1992), the Orion nebula (Perrin \& Sivan 1992), the
high-b dark nebula L1780 (Chlewicki \& Laureijs 1987; Mattila 1979),
and the starburst galaxy M82 (Perrin et al.\ 1995). The ERE was first
clearly recognized in the peculiar reflection nebula called Red
Rectangle by Schmidt et al.\ (1980).

2.\ The ERE peak wavelength varies from one environment to another,
and even within a given object, the peak shifts with distance from the
illuminating source. The density and hardness of the incident
radiation field appear to be the determining factors. The shortest
peak wavelength is seen for the ERE in the diffuse ISM ($\sim$610~nm)
in the case of the relatively weak interstellar radiation field, the
longest ($\sim$820~nm) is found on the Orion nebula bar adjacent to
the Trapezium stars, where the radiation density is several orders of
magnitude higher.

3.\ The photon conversion efficiency of the ERE, calculated on the
basis that all photons absorbed by dust in the 91.2-550~nm wavelength
range in a given system are absorbed by the ERE agent, with the
absorption of one UV/visible photon leading to the emission of at most
one ERE photon, has been determined to be near ($10 \pm 3$)\% in the
diffuse ISM (Gordon et al.\ 1998: Szomoru \& Guhathakurta 1998), and
it is at least this high in the Red Rectangle and not much lower in
the Orion nebula and a number of reflection nebulae.  This implies
that the ERE agent must absorb a significant fraction of the
UV/visible photons in the diffuse ISM and, simultaneously, possess an
exceedingly high intrinsic PL efficiency, $\gg$10\%. Since only part
of the energy of an absorbed photon is converted into ERE, the energy
conversion efficiency of the ERE agent is only about ($4 \pm 1$)\%.

4.\ No ERE is seen shortward of the wavelength of 540~nm. A sensitive
search for PL by dust in several reflection nebulae in the 400-500~nm
range, carried out by Rush \& Witt (1975), produced a null
result. While found in many dusty environments, ERE is essentially
absent in some, e.g.\ the Merope reflection nebula. The absence of ERE
in an otherwise normal dusty interstellar environment has not been
explained.

5.\ Among planetary nebulae, ERE was detected only in objects
currently thought to be carbon-rich, in the sense that C/O $> 1$
(Furton \& Witt 1992). This was thought to be a strong argument in
favor of the carbonaceous nature of the ERE carrier. This argument can
no longer be maintained in light of the ISO observations of C-rich
planetaries (Waters et al.\ 1998a) which show the presence of strong
spectral features of crystalline silicates in the mid-IR wavelength
region. In these planetaries, the oxygen-rich dust appears to be
present in the outer parts of the envelopes where material from
earlier mass-loss episodes resides. In NGC 7027, the only planetary in
which spatially resolved ERE observations have been made, the ERE is
found most strongly enhanced in the outermost envelopes (Furton \&
Witt 1990). In a similar way, ISO observations of the Red Rectangle by
Waters et al.\ (1998b), showing the presence of oxygen-rich dust there
as well, have greatly diminished the argument that the currently
C-rich stellar mass outflow implies a carbonaceous nature for the ERE
carrier in that object.  One important aspect of the ERE in planetary
nebulae is that the ERE band is seen superimposed upon the atomic
continuum only, i.e.\ there is no evidence of a scattered light
component. This would indicate that the dust particles in these
objects are sufficiently small to be in the Rayleigh limit, where
scattering becomes inefficient. ERE observations in planetary nebulae,
therefore, imply that the origin of ERE is connected with very small
grains.

6.\ In clumpy reflection nebulae, e.g.\ NGC 2023 and NGC 7023, ERE
appears strongly enhanced in filamentary structures, coincident with
surfaces of clumps seen in projection (Witt \& Malin 1989). It was
suggested that the ERE in these filaments becomes activated by the
exposure of carbonaceous materials to warm atomic hydrogen and UV
radiation in molecular hydrogen photodissociation zones (Furton \&
Witt 1993). However, high spatial resolution observations of H$_2$
vibrational fluorescence in NGC 2023 by Field et al.\ (1994, 1998)
showed that the correlation between ERE and H$_2$ structures is not
strong. In addition to some clearly correlated structures, there are
ERE filaments without corresponding H$_2$ filaments. This lack of
correlation is not explainable by optical depth effects, as is the
also observed case of H$_2$ filaments without ERE filaments.  An
analogous result was found when high-resolution H$_2$ observations of
NGC 7023 were compared to ERE structures in that nebula (Lemaire et
al.\ 1996).  Frequently, interstellar environments exhibiting ERE also
emit the near-IR emission bands (UIR bands) attributed to large
aromatic molecules, e.g.\ PAHs.  Often (e.g.\ in NGC 2023, 7023, 7027,
Orion nebula) both emission phenomena are seen in photodissociation
zones, but their intensities are poorly correlated.  In particular, in
the Orion bar, where both ERE and UIR bands have been observed, there
is no correlation between their respective intensities (Perrin \&
Sivan 1992). There is therefore little support to connect the emitters
of the ERE and of the UIR bands, except that they both occupy the same
general environments and that they respond to stellar UV illumination.

\section{Problems with previously proposed ERE candidates}

  Published models for materials causing the ERE have relied on
carbon-based substances, either solid-state carbonaceous solids or
carbon-based molecules. The former include hydrogenated amorphous
carbon (HAC) (Duley 1985), quenched carbonaceous composite (QCC)
(Sakata et al.\ 1992), and coal (Papoular et al.\ 1996); the latter
category includes polycyclic aromatic hydrocarbon (PAH) molecules
(d'Hendecourt et al.\ 1986) and C$_{60}$ (Webster 1993). Common
problems are the failure to match the observed ERE spectra with the
required PL efficiency.

  The most widely advocated model involves HAC (Duley et al.\ 1997;
Furton \& Witt 1993); it relies upon amorphous carbon, which is
already a component of several interstellar dust models.  The HAC PL
peak varies in wavelength in response to varying the conditions of
production and subsequent treatment; the absorption spectrum in the
blue and UV is smooth and does not introduce troublesome absorption
features unobserved in the ISM. It also explains the 3.4~$\micron$
C-H stretch feature observed in absorption along large lines of sight
in the diffuse ISM. The QCC model is similar in all of these respects.
Recent detailed laboratory studies of HAC (Robertson 1996: Rusli et
al.\ 1996, and refrences therein), however, have led to quantitative
results which cast doubts upon HAC as the ERE carrier. HAC is a highly
efficient photoluminescing material when its bandgap is large, near
4.4 eV. When illuminated by UV radiation, this HAC luminesces strongly
in the blue region of the spectrum, which makes it unsuitable as an
ERE analog. Narrowing the bandgap, e.g.\ by dehydrogenation, reduces
the PL efficiency of HAC exponentially, so that the HAC PL efficiency
drops to a few $10^{-4}$ of its maximum value when the bandgap is
small enough to yield PL emission in the wavelength region where ERE
is observed. HAC's principal problem, therefore, is its inability to
meet the spectral characteristics of the ERE and the ERE efficiency
simultaneously.

  The PAH model (d'Hendecourt et al.\ 1986) was a competing early
suggestion, which attributed the ERE in the Red Rectangle to
luminescence by large PAH molecules. However, even a large PAH like
the 13-ring hexabenzocoronene exhibits a sharply structured PL
spectrum peaking near 500~nm wavelength, and smaller PAHs luminesce
generally at still shorter wavelengths.  To obtain a broad,
unstructured band as observed in the ERE, d'Hendecourt et al.\ rely
upon a poorly studied process of intramolecular vibrational energy
randomization. The fact that ERE is seen only with peaks in the
610-820~nm wavelength range would require substantially larger PAHs
than hexabenzocoronene, with a total absence of smaller PAHs. A
further problem with all organic luminescing materials is that their
absorption spectra are highly structured as well, which should lead to
observable absorption bands in the blue and UV spectral regions of
highly reddened stars, which so far remains undetected. On the other
hand, the PAH model has had considerable success in explaining the
near-IR emission bands at 3.3, 6.2, 7.7, 8.6, and 11.3~$\micron$.

  The C$_{60}$ model (Webster 1993) can produce reasonable spectral
matches with the observed ERE, especially if mixtures of differently
sized fullerenes were used, but its fatal weakness is the measured PL
efficiency of $8.5 \times 10^{-4}$ for the C$_{60}$ fluorescence (Kim
et al.\ 1992). The ERE efficiency requirement is particularly
difficult to match for any molecular/organic luminescent material
which does not exhibit a broad continuous absorption spectrum covering
the 90-550~nm spectral range. The ERE requires the presence of a
material with a bandgap near 2 eV, which naturally absorbs photons of
higher energy and which converts part of the absorbed energy of these
photons into ERE photons with an efficiency of not less than ($10 \pm
3$)\%.

\section{Silicon nanoparticles as a potential ERE source}

  In this section, we shall discuss the PL characteristics of silicon
nanoparticles and how these meet the requirements posed by
astronomical observations of ERE. In the next section, we will explore
whether such particles can form under astronomical conditions and
whether cosmic abundances will permit the existence of sufficient
numbers of particles of this type.

  In order for materials to be efficient photoluminescing sources, two
requirements must be met: The electronic excitation resulting from the
absorption of a photon must be confined spatially, and possibilities
for non-radiative recombinations must be minimized. In organic
luminescent materials such as PAHs the confining units are individual
molecules, and photon yields can be as high as 99\% in some instances
(Krasovitskii \& Bolotin 1988). In the case of HAC, small aromatic
islands embedded in a sp$^3$-coordinated amorphous carbon matrix are
the absorbing entities. The much higher band gap energy of the
surrounding matrix confines the excited electrons to the aromatic
islands, where the PL can then occur (Robertson 1992, 1996). The
observation of ERE in planetary nebulae, which occurs in the absence
of visible scattering, confirms that the luminescing particles are
small compared to conventional interstellar grains.

  Silicon nanoparticles, which consist of crystalline silicon cores of
1-3~nm diameter, surrounded by a SiO$_{2}$ mantle, are known to be
remarkably efficient PL emitters in the 1.5-2.0~eV energy range (e.g.\
Wilson et al.\ 1993). The quantum confinement in such zero-dimensional
crystallites is responsible both for shifting the bandgap from a value
near 1.1~eV in bulk crystalline silicon to values of 2.0~eV and above
(Brus 1986, 1994; Delley \& Steigmeier 1993; Delerue et al.\ 1993) and
for greatly enhancing the PL efficiency (Efros \& Prigodin 1993) to
laboratory-measured values as high as 50\% at temperatures around 50~K
(Wilson et al.\ 1993). A series of recent investigations using
size-selected silicon nanostructures (Schuppler et al.\ 1994, 1995;
Ehbrecht et al.\ 1997; Lockwood et al.\ 1996) have established the
correlation between the structure size and the wavelength of peak PL
emission. According to these studies, silicon nanoparticles in the
1.5-5~nm diameter range luminesce strongly in the 600-850~nm
wavelength range. The reported maximum PL efficiencies ocurred for
sizes near 1.6~nm, corresponding to 664~nm wavelength, for
two-dimensional silicon nanostructures (Lockwood et al.\ 1996) and for
sizes near 3.9~nm, corresponding to 726~nm wavelength, for
zero-dimensional nanoparticles (Ehbrecht et al.\ 1997), and both
studies reported a rapid drop in PL efficiency when going to both
larger and smaller size parameters.

\vspace*{0.1in}
\begin{center}
\epsscale{0.50}
\plotone{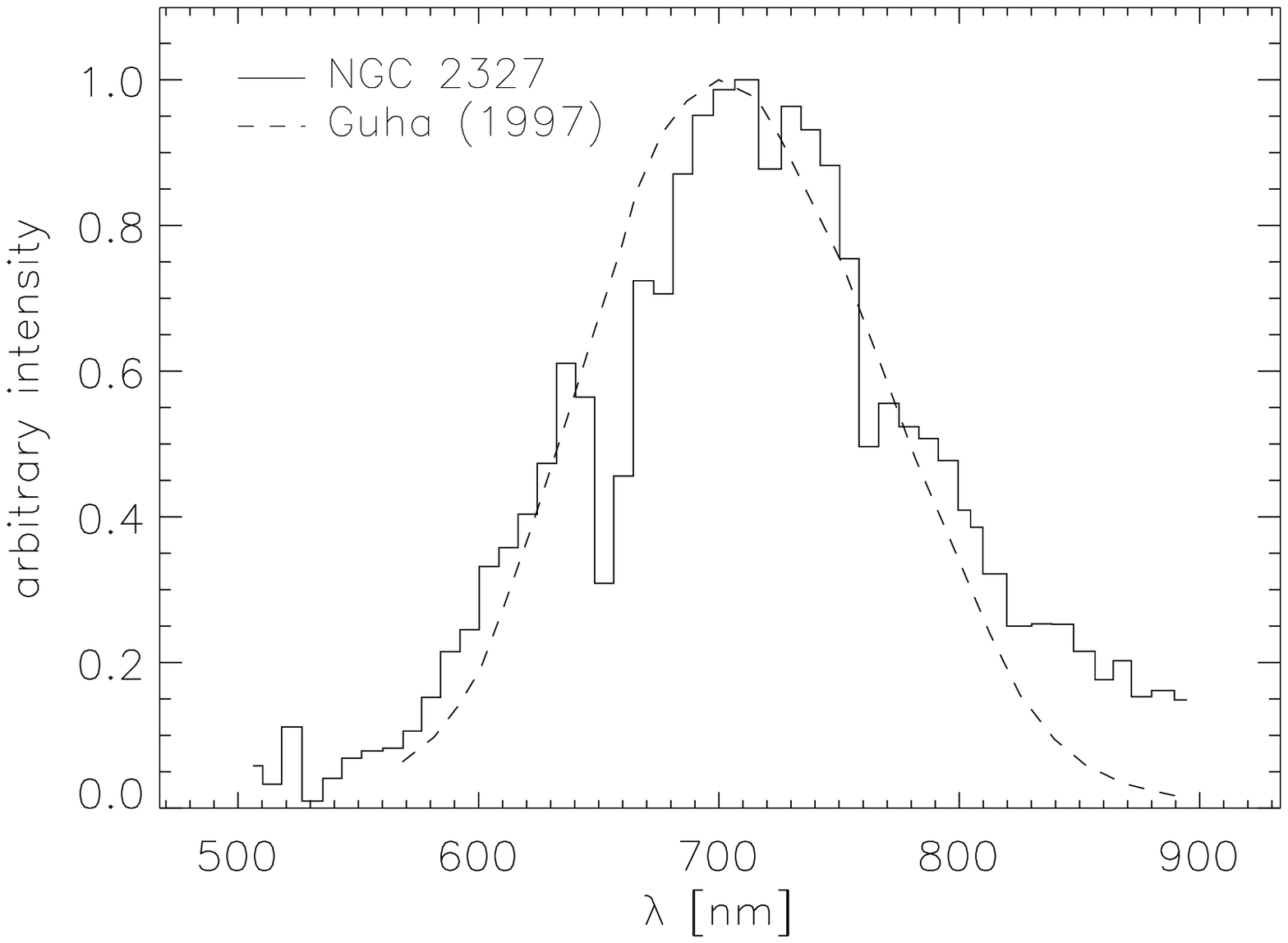}
\end{center}
Fig.\ 1 -- The observed ERE spectrum of NGC 2327 (Witt 1988) is plotted
along with the photoluminescence of film B at room temperature from
Guha (1997).  Film B had a porosity of 70\% which corresponds to a
particle size of 4~nm.
\vspace*{0.2in}

  We conclude from the published experimental results that silicon
nanoparticles luminesce extremely efficiently (up to 50\%) in the
600-850~nm range when they occur in a very limited size range from
1.5-5.0~nm, and not otherwise. The width of the PL band depends on the
width of the size distribution present, and the wavelength of maximum
PL intensity is determined by the dominant size within the
distribution. The published PL spectra of silicon nanoparticles match
those observed in ERE sources extremely closely (see
Fig.~1). Environmental effects, such as vaporization of the smaller
particles in intense radiation fields, will lead to a gradual shift of
maximum PL toward longer wavelength, as observed. On the other hand,
the diffuse ISM with the lowest radiation densities will allow the
existence of the smaller particles as well, with a resultant maximum
efficiency and a very broad PL spectrum.

\section{Silicon nanoparticles as a component of interstellar dust}

  Any interstellar dust component candidate must provide positive
answers to two fundamental questions. Can the grains form naturally in
an astronomical environment? Is the required total grain mass
consistent with the cosmic elemental abundances of the constituents?

  Silicon monoxide (SiO) is one of the most abundant and most strongly
bonded molecules in the outflows from oxygen-rich stellar sources;
virtually all silicon in the gas phase is initially locked up in this
form (Gail \& Sedlmayr 1986). The nucleation of silicates is thought
to be precipitated by the initial formation of SiO clusters (Nuth
1996), a conclusion supported by laboratory evidence regarding the
condensation of SiO (Nuth \& Donn 1982), which showed the condensate
to contain elemental silicon plus silicon oxide SiO$_x$ with $x \sim
1.5$.  This can be understood by the fact that SiO$_2$ is the
energetically preferred form of silicon oxide in the solid phase,
while SiO is preferred in the gas phase.  The competition for
additional oxygen atoms in the condensing cluster will leave half of
the silicon atoms ultimately without a partner, if the change of state
can complete itself. Since this condensation occurs in a
high-temperature environment ($\sim$500-1000~K), we postulate that the
resulting annealing of the condensing SiO clusters will lead to a
separation of the elemental silicon into a core and the SiO$_2$ into a
mantle, to form the basic structure of silicon nanocrystals with
oxygen passivation (Littau et al.\ 1993), i.e.\ in which the dangling
bonds at the surface of the silicon core are connected to oxygen atoms
in the SiO$_2$ mantle. As we will show, only 5\% of the total dust
mass needs to remain in the form of silicon nanoparticles; the
overwhelming fraction of the condensing particles can, therefore,
remain involved in the condensation of various silicates and metal
oxides, as supported by astronomical observations. We conclude,
therefore, that silicon nanoparticles are a product of the initial
dust formation process occurring in one of several oxygen-rich outflow
environments, e.g.\ M-type supergiants, WN stars, type II supernovae,
and AGB stars. In addition to producing the ERE, these silicon
nanoparticles will be subject to temperature fluctuations, resulting
from absorptions of individual UV photons, and thus contribute to the
near- and mid-IR thermal background radiation received from the
diffuse ISM.  The temperature increases expected from the absorption
of individual 10 eV photons (the absorption peak of the Si cores is
near 125 nm) by Si nanoparticles in the 1.5 - 5 nm size
range are 170 K - 60 K (Purcell 1976).  These are upper limits because
part of the absorbed energy reemerges as ERE photons, and the actual
particle mass is larger than assumed here by the amount of SiO$_2$
contained in the mantle. The Si nanoparticles required for efficient
ERE production are therefore not hot enough to cause significant
emission in the 9.5 $\micron$ Si-O band from their SiO$_2$
mantles. That would require temperatures in excess of 230 K. Our
proposal is therefore not in conflict with the conclusion of Desert et
al.\ (1986) that the smallest interstellar grains are most likely
carbonaceous in nature.  However, the SiO$_2$ mantles should
contribute to the Si-O features at 9.5 $\micron$ and 20 $\micron$ in
absorption, which are generally referred to as the interstellar
silicate features (Whittet 1992). To conclude, in contrast to some
other ERE candidates, silicon nanoparticles are not expected to
produce spectral features which are not already observed in the
diffuse ISM.

  The question of abundance is of particular importance, especially in
view of the suggested reductions in the interstellar heavy element
abundances relative to hydrogen (Snow \& Witt 1996), compared to
formerly used solar abundances. Silicon absorbs strongly at energies
above 3~eV, with an average absorption coefficient of $1.5 \times
10^6$~cm$^{-1}$ throughout the UV (Landolt-Bornstein 1982). With a
density of 2.42~g~cm$^{-3}$ for crystalline silicon, we derive a
cross-section per Si-atom of $2.9 \times
10^{-17}$~cm$^2$~Si$^{-1}$. Silicon nanoparticles luminesce with an
intrinsic efficiency of about 50\% (Wilson et al.\ 1993). Coupled with
the observed lower limit to the ERE photon conversion efficiency of
($10 \pm 3$)\% (Gordon et al.\ 1998), this suggests that silicon
nanoparticles are responsible for 20\% of the UV/visible absorption in
the ISM. Hence, along a typical interstellar sightline of 1~kpc with
an extinction of 3~mag~kpc$^{-1}$ in the UV and a typical dust albedo
of 0.5, silicon nanoparticles contribute about 0.3 magnitudes of
UV/visible absorption. This absorption and the cross-section per atom
leads to a density of silicon atoms required for nanoparticles of $3
\times 10^{-6}$~Si~cm$^{-3}$, while the hydrogen density for the
chosen environment is about 1~H~cm$^{-3}$. A similar number of silicon
atoms and twice that number of oxygen atoms will be needed for the
SiO$_2$ mantles.  The total amount of silicon in dust is $1.7 \times
10^{-5}$ relative to hydrogen; thus, by number only 36\% of the
interstellar silicon atoms present in interstellar dust need to be
present in the form of silicon nanoparticles in order for the ERE
observations to be explained.  The total mass of the silicon
nanoparticles with mantles, when compared with the total mass of
depleted interstellar atoms (Snow \& Witt 1996) amounts to less than
5\%.

\section{Summary}

  We have examined the observed characteristics of the ERE, a
photoluminescence phenomenon associated with interstellar dust and
seen in a wide variety of astronomical environments as excess
radiation in the 600-850~nm wavelength range. In the diffuse ISM,
about 10\% of absorbed UV/visible photons contribute to the ERE. The
ERE carrier must be a cosmically abundant material exhibiting highly
efficient photoluminescene properties. Silicon nanoparticles
containing a few hundred silicon atoms each, surrounded by a SiO$_2$
shell, match the constraints regarding the spectrum and efficiency of
the ERE.  Interstellar silicon abundances easily suffice to provide
the needed material quantities, and the nucleation of SiO in
oxygen-rich stellar outflows provides a likely source for their
formation. We suggest that silicon nanoparticles are an abundant
component of the interstellar dust spectrum.

\acknowledgements

   ANW acknowledges stimulating and fruitful exchanges of ideas with
Drs.\ H.- P.\ Gail, S.\ Guha, F.\ Huisken, K.\ D.\ Kolenbrander, D.\
Lockwood, and E.\ Werwa.  We also acknowledge constructive comments
from an anonymous referee.  This work was supported by grants from
NASA to The University of Toledo.

NOTE: After submission of this paper, we received an advance copy of a
paper by Ledoux et al. (1998), which arrives at conclusions similar to
ours.

\end{document}